\documentclass[twocolumn,  tighten,  twocolappendix]{aastex631}

\shorttitle{Impact of Galaxy Mergers on Stellar Kinematics of Early-type Galaxies}
\shortauthors{Yoon et al. (2022)}

\begin{document}

\title{Evidence for Impact of Galaxy Mergers on Stellar Kinematics of Early-type Galaxies}

\email{yyoon@kias.re.kr}

\author[0000-0003-0134-8968]{Yongmin Yoon}
\affiliation{School of Physics, Korea Institute for Advanced Study (KIAS), 85 Hoegiro, Dongdaemun-gu, Seoul, 02455, Republic of Korea}

\author[0000-0001-9521-6397]{Changbom Park}
\affiliation{School of Physics, Korea Institute for Advanced Study (KIAS), 85 Hoegiro, Dongdaemun-gu, Seoul, 02455, Republic of Korea}

\author[0000-0002-3043-2555]{Haeun Chung}
\affiliation{University of Arizona, Steward Observatory, 933 N Cherry Avenue, Tucson, AZ 85721, USA}

\author{Richard R. Lane}
\affiliation{Centro de Investigación en Astronomía, Universidad Bernardo O'Higgins, Avenida Viel 1497, Santiago, Chile}

\begin{abstract}
We provide observational evidence that galaxy mergers significantly affect stellar kinematics of early-type galaxies (ETGs) such as specific stellar angular momentum within the half-light radius ($\lambda_{R_e}$) and kinematic misalignment ($\psi_\mathrm{mis}$), using MaNGA integral field unit spectroscopic data that are in the Stripe 82 region of the Sloan Digital Sky Survey. In this study, tidal features around ETGs, which are detected in deep coadded images, are used as direct evidence for mergers that occurred recently. In the case of ETGs that do not have dust lanes, $\lambda_{R_e}$ is lower in ETGs with tidal features than in those without tidal features (median $\lambda_{R_e}$: $0.21$ versus $0.39$) in all stellar mass and S{\'e}rsic index ranges except the most massive bin, so that the fraction of ETGs with tidal features in slow rotators is more than twice as large as that in fast rotators ($42\%$ versus $18\%$). Moreover, ETGs with tidal features have larger $\psi_\mathrm{mis}$ than those without tidal features (mean $\psi_\mathrm{mis}$: $28\degr$ versus $15\degr$). By contrast, ETGs with dust lanes are fast rotators, and ETGs with both dust lanes and tidal features have the highest $\lambda_{R_e}$ (median $\lambda_{R_e}$: $0.59$) among all ETG categories. In addition, ETGs with dust lanes have small $\psi_\mathrm{mis}$ regardless of the existence of tidal features ($\psi_\mathrm{mis}<7.5\degr$). Our results can be explained if mergers with different gas fractions generate merger remnants that have different kinematic properties.
\end{abstract}

\keywords{Early-type galaxies (429) --- Galaxy kinematics (602) --- Galaxy mergers (608) --- Galaxy rotation (618)  --- Galaxy properties (615) --- Tidal tails (1701)}

\section{Introduction}\label{sec:intro}

Early-type galaxies (ETGs) are evolved systems in which most of stellar populations are old. The star formation activities in ETGs have been quenched due to depletion of available cold gas in the systems. Consequently, ETGs show red colors ($g-r\gtrsim0.7$) in optical bands \citep{Park2005,Gallazzi2006,Graves2009,Choi2010,Schawinski2014,YP2020}. Another photometrically discernible property of ETGs is that they have more centrally concentrated light distributions than late-type galaxies \citep{Park2005,Choi2010}. ETGs also have relatively smooth structures or simple shapes compared with late-type galaxies that usually have clearly distinguishable components (red bulges and blue disks) with spiral arms and sometimes individual blobs of highly star forming regions \citep{Nair2010}.

The advent of surveys based on integral field unit (IFU) spectroscopy\footnote{For example, the SAURON project \citep{Bacon2001}, the MASSIVE survey \citep{Ma2014}, the DiskMass survey \citep{Bershady2010}, the ATLAS$^\mathrm{3D}$ project \citep{Cappellari2011}, the CALIFA survey \citep{Sanchez2012,Sanchez2016}, and the SAMI survey \citep{Croom2021}.} enables us to understand diverse kinematic properties of ETGs that could not be obtained simply through photometry. Thus, IFU surveys have opened up new perspectives on ETGs. For example, it is found that a significant number of ETGs, even of those with apparently round and nondisky shapes, have substantial (fast) rotating stellar components that are similar to disks in late-type galaxies \citep{Cappellari2016,Graham2018}. Hence, many recent studies classified ETGs according to kinematic properties of ETGs such as the degree of rotation (or angular momentum) that stellar components of ETGs have (e.g., slow and fast rotators; \citealt{Emsellem2007,Jesseit2009,Emsellem2011,Khochfar2011,Cappellari2016,Penoyre2017,Graham2018}), and suggested that it is a physically meaningful classification for ETGs to replace the traditional separation into ellipticals and lenticulars that is subject to a strong dependence on line-of-sight inclinations \citep{Emsellem2007,Cappellari2011,Cappellari2016}.

The formation and growth of massive ETGs are known to be closely related to galaxy mergers. Mergers can produce quiescent remnants that do not form young stellar populations in the end, due to exhaustion of available cold gas through intense star formation \citep{Hernquist1989,Mihos1996,Springel2005} and perhaps feedback effects from active galactic nuclei (AGNs; \citealt{Hopkins2005,Hopkins2008b,Springel2005}) during the merger processes. Mergers are also capable of making concentrated stellar light distributions in post-merger galaxies \citep{Barnes1988,Naab2006,Hilz2013}. Moreover, minor mergers with little gas have a significant effect on the size growth of ETGs \citep{Bernardi2011,Oogi2013,Yoon2017}. That said, there is also strong evidence that the early morphological type of galaxies can result from close interactions with neighboring ETGs \citep{Park2008,Park2009}, which has been confirmed by numerical simulations of galaxy--galaxy interactions \citep{Hwang2015,Hwang2018}.

It is natural to expect that galaxy mergers also influence stellar kinematics of ETGs, given the fact that specific stellar angular momentum tends to be lower for more massive ETGs \citep{Emsellem2007,Graham2018} that form and grow more predominantly through mergers \citep{DeLucia2006,DeLucia2007,Yoon2017}. There have been studies based on numerical simulations to reveal how galaxy mergers influence internal stellar kinematic properties of merger remnants. For example, \citet{Choi2017} showed that mergers statistically contribute to the lowering of rotation speeds especially in massive galaxies, and frequent minor mergers have strong cumulative effects on kinematics of merger remnants. \citet{Bois2011} and \citet{Jesseit2009} claimed that mass ratios and merger orbits of progenitors are crucial factors for the kinematics of merger remnants. Some studies \citep{Cox2006,Hoffman2010,Naab2014,Penoyre2017} argued that gas fractions in progenitors are also an important factor that affects the kinematics of merger remnants. It has been also shown through numerical simulations that the spin angular momentum of galaxies drops through prograde galaxy--galaxy encounters \citep{Hwang2021}.  

Compared with simulation works, it is relatively difficult to conduct direct observational studies about the effect of mergers on properties of ETGs, since we only observe a snapshot of the universe, not the stream of cosmic time. However, mergers between galaxies leave tidal features around the remnants that are detectable in deep images \citep{Toomre1972,Quinn1984,Barnes1988,Hernquist1992,Feldmann2008}. Therefore, tidal features such as tidal tails, streams, and shells are considered the most direct observational evidence for recent mergers, through which it is possible to investigate the influence of mergers on photometric \citep{Schweizer1992,Tal2009,Schawinski2010,Kaviraj2011,Hong2015,YL2020} and kinematic \citep{Krajnovic2011,Duc2015,Oh2016} properties of ETGs. For instance, \citet{Hong2015} suggested that luminous AGNs in ETGs are related to galaxy mergers based on the discovery that almost half of luminous AGN hosts have tidal features. \citet{YL2020} discovered direct evidence that compact young ETGs especially with blue cores and ETGs with dust lanes are involved in recent mergers, by investigating how the fraction of ETGs having tidal features depends on the age and internal structure of ETGs.

Here, we study the influence of recent mergers on stellar kinematics of ETGs such as stellar angular momentum and kinematic misalignments, using ETGs in the Mapping Nearby Galaxies at Apache Point Observatory (MaNGA; \citealt{Bundy2015,Drory2015,Yan2016,Wake2017}) IFU data that are in the Stripe 82 region of the Sloan Digital Sky Survey (SDSS). The unprecedented number of uniformly observed galaxies from the MaNGA survey combined with deep coadded images of the Stripe 82 region leads us to better understand the role of recent mergers on stellar kinematics of ETGs.

 In this study, we use \emph{H$_0=70$} km s$^{-1}$ Mpc$^{-1}$, $\Omega_{\Lambda}=0.7$, and $\Omega_\mathrm{m}=0.3$ as cosmological parameters. 
\\

\section{Sample and Analysis}\label{sec:sample}
\subsection{SDSS-IV MaNGA}\label{sec:manga}
The MaNGA IFU survey is a project of the fourth generation of SDSS \citep{Blanton2017} using the ARC 2.5 m telescope \citep{Gunn2006}. The spectrograph of the MaNGA project is the same as that of the Baryonic Oscillation Spectroscopic Survey \citep{Smee2013}. Its wavelength coverage is 3600 -- 10300\,\AA, in which the midrange spectral resolution corresponds to $\sim2000$. The observation was conducted with 17 fiber-bundled IFUs that are assigned within a $3\degr$ field of view. The sizes of the fiber-bundled IFUs are from $12\arcsec$ to $32\arcsec$ depending on the number of fibers. The selection of target galaxies for the MaNGA project was based on $i$-band absolute magnitude and redshift (and near-ultraviolet$- i$ color for a small portion of targets). The target galaxies are evenly distributed in the color-magnitude space and have uniform spectroscopic coverage up to a 1.5 or 2.5 half-light radius along the major axis ($R_e$). More details about the selection of the target galaxies are in \citet{Wake2017}.

In this study, we use MaNGA data of the internal data release version MPL-11, which is equivalent to the final release version (Data Release 17) that contains all observational data of this project. A small portion ($\sim 1\%$) of MaNGA data were obtained through repeated observations of the same galaxy. In those cases, we selected one observation that has the largest IFU size among the duplicate ones. If the IFU sizes of the duplicate observations are identical, we chose the data with the highest blue channel signal-to-noise ratio (S/N). 
\\

\subsection{Deconvolution of IFU Data}\label{sec:decon}

The atmospheric seeing and the aberration from the instrument optics that deteriorate spatial resolutions of observational data are inevitable factors in data obtained from ground-based telescopes. The seeing effect is more severe in fiber-based IFU data due to large physical gaps between sampling elements (fibers). Thus, if such an effect is minimized, we can improve spatial resolutions of IFU data and obtain more robust kinematic information for galaxies, particularly in the central part of galaxies. To mitigate the seeing effect, we deconvolved MaNGA IFU data using the Lucy--Richardson (LR) algorithm \citep{Richardson1972, Lucy1974} as in \citet{Chung2020} and \citet{Yoon2021}. The LR algorithm is an iterative procedure that recovers an original image that has been convolved by a point spread function (PSF). This algorithm is advantageous in deconvolution of very large data, since it requires only a few parameters to perform. Here, we only concisely describe the implementation of the LR algorithm on MaNGA IFU data. Details on the algorithm and its application to MaNGA data cubes are contained in \citet{Chung2020} (see also \citealt{Yoon2021}).

The simple equation form for the LR algorithm is
\begin{equation}
u^{n+1}=u^n\cdot \Bigg[\bigg(\frac{d}{u^n\otimes p}\bigg)\otimes p\Bigg],
\label{eq:deconv}
\end{equation}
in which $u^n$ is $n^\mathrm{th}$ estimation of the maximum likelihood solution and $d$ denotes an original PSF-convolved image ($u^0=d$). The term $p$ indicates a two-dimensional (2D) PSF and $\otimes$ means 2D convolution. We applied the LR algorithm individually to the 2D image slice of each wavelength bin in MaNGA data cubes. The FWHM of the PSF at each wavelength bin was determined by interpolation from the linear function fitted to the FWHM values at the $g$, $r$, $i$, and $z$ bands, which are recorded in MaNGA data cubes. We used a 2D Gaussian function with the determined FWHM as a functional form of the PSF to deconvolve all of the pixels in a 2D image slice. We note that PSFs of MaNGA data can be well described by a single 2D Gaussian function and the FWHM varies less than $10\%$ across a given IFU \citep{Law2015,Law2016}.
\citet{Chung2020} tested how the number of iterations ($N_\mathrm{iter}$) in the LR algorithm affects the quality of deconvolution and found that beyond  $N_\mathrm{iter}=20$, the quality of deconvolution is not significantly improved, while additional artifacts can be generated in the image with enhanced noise. Therefore, we fixed $N_\mathrm{iter}$ to 20, which is the optimum iteration number in the deconvolution process found in \citet{Chung2020}.

Intensive tests of the application of the LR algorithm to mock IFU data in \citet{Chung2020} demonstrate that the deconvolution process allows us to effectively recover the true stellar kinematics of galaxies. For instance, the test showed that the luminosity-weighted stellar angular momentum (Equation \ref{eq:lambda}) can be recovered with the underestimation of less than $\sim5\%$ in most cases in which the deconvolution process is applied. That said, the underestimation can rise up to $\sim20\%$--$30\%$ when the deconvolution is not applied. 
\\

\begin{figure*}
\includegraphics[width=\linewidth]{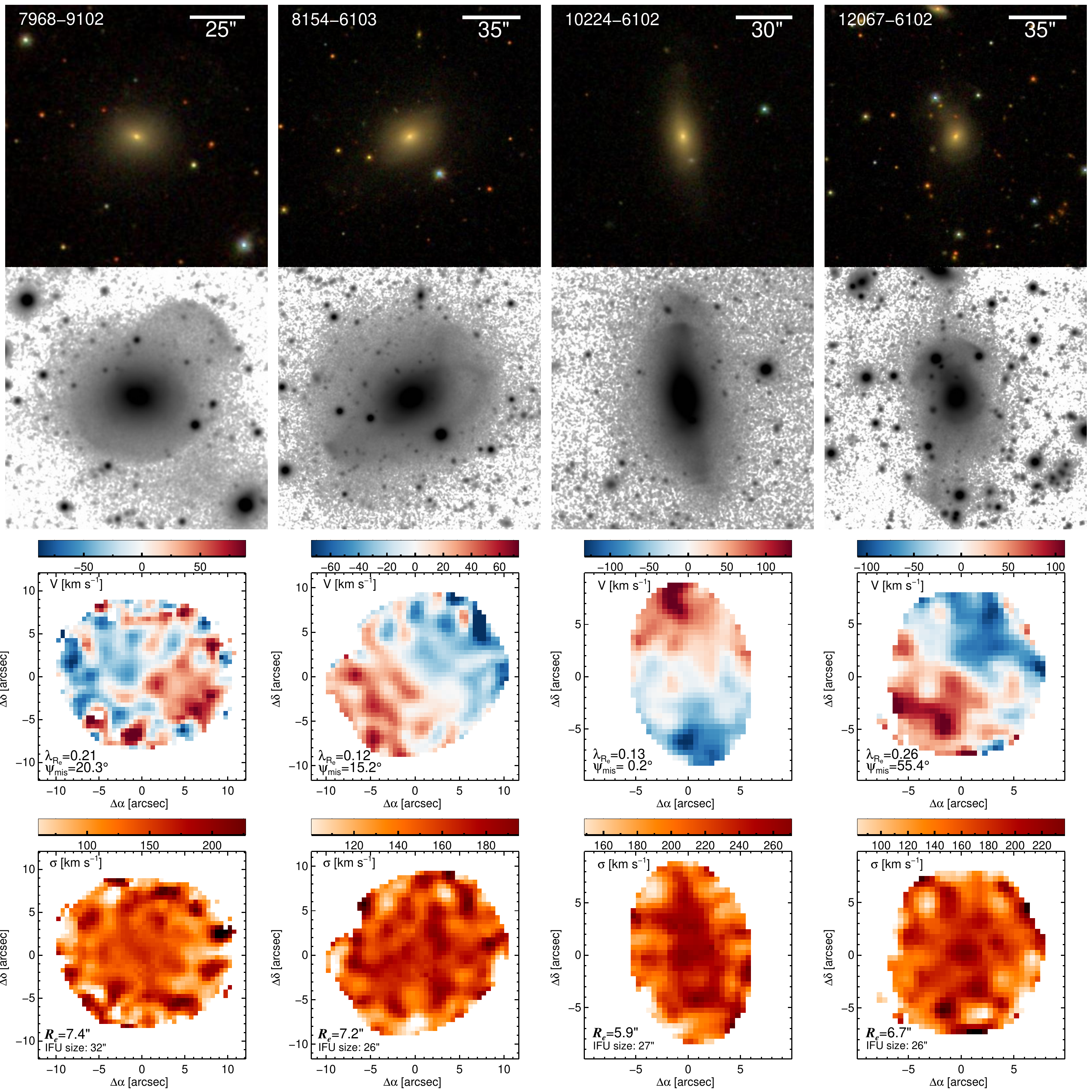}  
\centering
\caption{Examples of ETGs with tidal features. The first row: single-epoch color images in SDSS. The horizontal bars in the color images are scales of the images. The Plate ID (e.g., 7968) and IFU design ID (e.g., 9102) are written in the color images. The second row: deep coadded images. The image scale of each deep image is the same as that of the color image in the first row. The third row: extracted 2D line-of-sight velocity maps within $1.5R_e$. The $\lambda_{R_e}$ and $\psi_\mathrm{mis}$ of each galaxy are shown in the lower left corner of each panel in the third row. The fourth row: extracted 2D line-of-sight velocity dispersion maps within $1.5R_e$. The IFU size and $R_e$ of each galaxy are shown in the left corner of each panel in the fourth row. See the color bars above the panels in the third and fourth rows for the color-coded velocity or velocity dispersion scales. The terms $\Delta\alpha$ and $\Delta\delta$ are relative R.A. and decl., respectively. 
\label{fig:ex_1}}
\end{figure*} 

\begin{figure*}
\includegraphics[width=\linewidth]{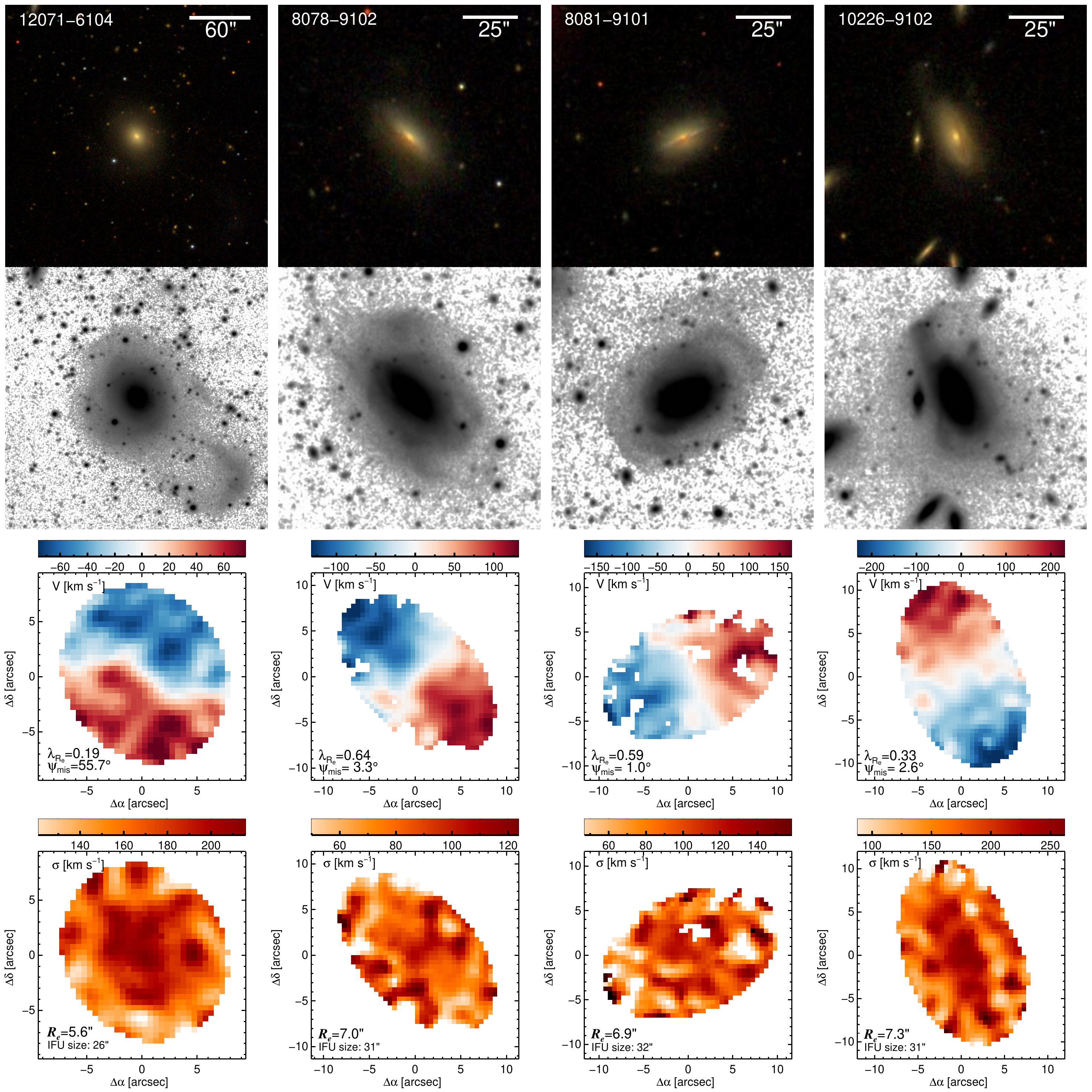}  
\centering
\caption{Examples of ETGs with tidal features. Descriptions about the figure are identical to those in Figure \ref{fig:ex_1}. The second, third, and forth columns represent ETGs that have both dust lanes and tidal features.
\label{fig:ex_2}}
\end{figure*} 

\begin{figure*}
\includegraphics[width=\linewidth]{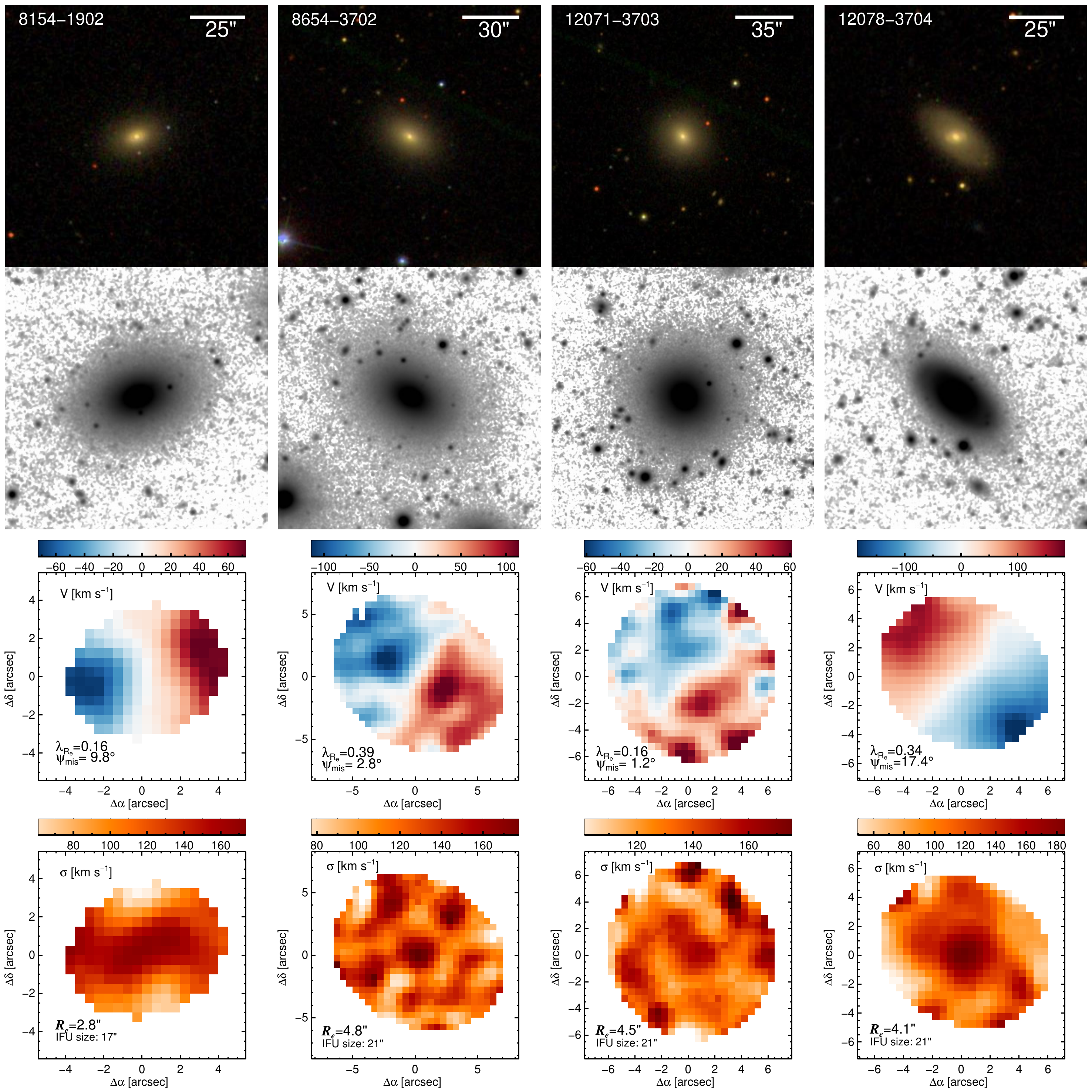}  
\centering
\caption{Examples of normal ETGs without tidal features. Descriptions about the figure are identical to those in Figure \ref{fig:ex_1}.
\label{fig:ex_3}}
\end{figure*} 

\subsection{Extracting Stellar Kinematics}\label{sec:extract}
The method used in \citet{Yoon2021} was utilized to extract stellar kinematics from MaNGA data such as line-of-sight velocities and velocity dispersions. We utilized the code Penalized Pixel-Fitting \citep[pPXF;][]{Cappellari2004,Cappellari2017}. pPXF uses the maximum likelihood formalism and conducts full spectrum fitting on galaxy spectra to extract stellar kinematics and stellar populations. For input model templates to run pPXF, we used MILES single stellar population models \citep{Sanchez2006,Vazdekis2010,Falcon2011} with the initial mass function of \citet{Chabrier2003} and the Padova+00 isochrone \citep{Girardi2000}. In total, 60 model templates were used: the model templates have 10 age steps from 0.07 Gyr to 12.59 Gyr and six metallicity values in the range $-1.71\leq[M/H]\leq0.22$.\footnote{The upper limit of $[M/H]=0.22$ for the model templates can be low for the most massive ETGs that are expected to have high metallicity values. However, the use of alternative MILES models based on the BaSTI (a Bag of Stellar Tracks and Isochrones) isochrone \citep{Hidalgo2018} that reaches a higher metallicity of $[M/H]=0.4$ does not change the extracted kinematics of the 13 most massive ETGs with $\log(M_\mathrm{star}/M_\odot)>11.2$ in our final sample: the maximum difference in $\lambda_{R_e}$ is only 0.006.} As in \citet{Belfiore2019}, eighth-order additive Legendre polynomials were added to the templates in order to improve the quality of the extracted stellar kinematics \citep{Emsellem2004}.

Before the fitting process, we masked pixels around known emission lines and bad pixels (pixels with dead fibers, low coverage depth, or contamination from foreground stars, etc.) flagged in the MaNGA data reduction pipeline \citep{Law2016}. The model spectra were convolved with a Gaussian function to match the resolution of MaNGA spectra, and MaNGA IFU spectra were shifted to the rest frame before extracting stellar kinematics following the method described in \citet{Cappellari2017}. The fitting was conducted in a wavelength range from 3700 to 7400\AA, considering the fact that the wavelength coverage of the model spectra is 3540 -- 7410\AA. We show examples of extracted 2D line-of-sight velocity and velocity dispersion maps in the third and fourth rows of Figures \ref{fig:ex_1}, \ref{fig:ex_2}, and \ref{fig:ex_3}.

We measure the dimensionless spin parameter $\lambda_R$ (luminosity-weighted specific stellar angular momentum) as in \citet{Emsellem2007}:
\begin{equation}
\lambda_R\equiv\frac{\langle R \, \vert V \vert \rangle}{\langle R \, \sqrt{V^2+\sigma^2} \rangle }=\frac{\sum^{N}_{i=1}F_i\,R_i\, \vert V_i \vert}{\sum^{N}_{i=1}F_i\,R_i \, \sqrt{V_i^2+\sigma_i^2}},
\label{eq:lambda}
\end{equation}
where $F_i$ is the flux of the $i$th bin and $R_i$ is the circular radial distance from the center to the $i$th bin (here we used $F_i$ of $r$-band images decovolved in the same way as in IFU data). $V_i$ and $\sigma_i$ are the line-of-sight velocity and velocity dispersion of the $i$th bin, respectively. The summation is conducted over $N$ pixels within the photometric ellipse. Spin parameter $\lambda_R$ is normalized by $\sqrt{V_i^2+\sigma_i^2}$ that is a proxy for mass, and goes to unity when the system is rotation dominated and becomes 0 when the system is fully supported by random motion. In the calculation of $\lambda_R$, we only used spaxels with median S/N $\ge5$. Spaxels with $\sigma<40$ km s$^{-1}$ were excluded in the calculation, since such small $\sigma$ values are uncertain due to the instrumental resolution limit \citep{Penny2016,Lee2018}.\footnote{In order to examine whether the prescription that excludes spaxels with low $\sigma$ changes $\lambda_R$ values significantly, we conducted a test in which we fixed $\sigma$ values to 40 km s$^{-1}$ for spaxels with $\sigma<40$ km s$^{-1}$ and included them in the calculation of $\lambda_R$. In this test, we found that setting the lower limit of $\sigma=40$ km s$^{-1}$ makes $\lambda_{R_e}$ of our final ETG sample increase only by 0.007 on average, compared to the original process that excludes spaxels with $\sigma<40$ km s$^{-1}$. Furthermore, $97\%$ of $\lambda_{R_e}$ values of our ETG sample do not increase by more than 0.05 by setting the lower limit in $\sigma$. Therefore, excluding spaxels with low $\sigma$ values does not change our results significantly.} We also excluded spurious spaxels with $\vert V \vert\ge500$ km s$^{-1}$ in the calculation of $\lambda_R$. 

Spin parameter $\lambda_R$ is a widely used proxy for the stellar angular momentum of galaxies \citep{Jesseit2009,Emsellem2011,Fogarty2015,Cappellari2016,Oh2016,Choi2017,Graham2018}. In particular, many previous works based on observational data have used $\lambda_R$ within $R_e$ ($\lambda_{R_e}$) for statistical studies of galaxy stellar angular momentum. We note that $R_e$, ellipticities ($\varepsilon$), and photometric position angles ($\psi_\mathrm{phot}$) of galaxies were derived based on the $r$-band elliptical Petrosian flux.\footnote{Note that $\varepsilon$ and $\psi_\mathrm{phot}$ are calculated at $90\%$ light radius.} Stellar masses ($M_\mathrm{star}$) of galaxies used in this study were also calculated from the elliptical Petrosian fluxes. These parameters are from the NASA-Sloan Atlas (NSA) catalog, which is a base catalog for the selection of the galaxies in the MaNGA survey \citep{Wake2017}. 

Parameter $\lambda_R$ is not changed much for a wide range of viewing angles, unless the angle is close to face-on \citep{Emsellem2007,Jesseit2009,Bois2011}. This is due to the fact that $\langle V \rangle$ and $\langle \sigma \rangle$ simultaneously decrease as the inclination declines so that the ratio between them is not substantially changed \citep{Jesseit2009}. Thus, $\lambda_R$ is a robust observational indicator for the intrinsic angular momentum in most galaxies. 

The kinematic position angles for galaxies ($\psi_\mathrm{kine}$) were measured using the method in Appendix C of \citet{Krajnovic2006}. For an observed velocity map $V(x,y)$, we generated multiple model velocity maps $V'(x,y)$ whose $x$-axes are along various angles. Specifically, $V'(x,y)$ was computed by averaging observed velocity values in the four quadrants: $V'(x,y)=[V(x,y)+V(x,-y)-V(-x,y)-V(-x,-y)]/4$. Then, the position angle of a model velocity map that has the minimum $\chi^2$ difference between the model and the observed one was determined as $\psi_\mathrm{kine}$ of the galaxy. In this study, the kinematic misalignment ($\psi_\mathrm{mis}$) is defined as 
\begin{equation}
\sin\psi_\mathrm{mis}=\vert \sin(\psi_\mathrm{kine} - \psi_\mathrm{phot}) \vert.
\label{eq:miss}
\end{equation}
Our results of this study do not change even if we use $\psi_\mathrm{phot}$ at $50\%$ light radius, which is likely to be less affected by surrounding tidal features, in the calculation of $\psi_\mathrm{mis}$, instead of $\psi_\mathrm{phot}$ at $90\%$ light radius. Indeed, $\psi_\mathrm{phot}$ at $90\%$ light radius, which we adopted in this study, is a bit better aligned with $\psi_\mathrm{kine}$ than $\psi_\mathrm{phot}$ at $50\%$ light radius, even for ETGs with tidal features.\footnote{The mean $\psi_\mathrm{mis}$ for our final sample is $16.7\degr$ when $\psi_\mathrm{phot}$ at $90\%$ is used, while it is $19.9\degr$ when $\psi_\mathrm{phot}$ at $50\%$ is used.}
\\

\subsection{Selection of ETG Samples with Fine Qualities in Stripe 82}\label{sec:sam}

In this study, we used MaNGA galaxies in the Stripe 82 region of SDSS. The Stripe 82 region covering $\sim300\,\mathrm{deg}^{2}$ was scanned $\sim70$--$90$ times in the imaging survey. The detection limit of coadded images of Stripe 82 is $\sim2$ mag deeper than single-epoch images of SDSS \citep{Jiang2014}. Therefore, faint tidal features of ETGs that are difficult to detect in the SDSS single-epoch images can be discovered in the coadded images of Stripe 82 \citep{Kaviraj2010,Schawinski2010,Hong2015,YL2020}.

The number of MaNGA galaxies in the coadded images of Stripe 82 is 699. We only used galaxies in the redshift range $z <0.055$. Galaxies at higher redshifts were excluded due to their small angular sizes and the effect of the cosmological surface brightness dimming (see Equation 6 in \citealt{YP2020}) that make the detection of tidal features difficult. We only used galaxies with $M_\mathrm{star}\ge10^{9.65}\,M_{\odot}$, since no tidal feature was detected at $M_\mathrm{star}<10^{9.65}\,M_{\odot}$ in the final sample. Applying the redshift and stellar mass cuts, the number of galaxies is 525. 

We classified ETGs from the sample by visually inspecting combined color images of $g$, $r$, and $i$ bands in SDSS (see the first row of Figures \ref{fig:ex_1}, \ref{fig:ex_2}, and \ref{fig:ex_3}) as in \citet{Yoon2021}, in which we visually classified T-types of galaxies for more than 2000 MaNGA galaxies. See \citet{Yoon2021} for details about the morphology classification. By visual inspection, 254 galaxies were classified as ETGs among the 525 galaxies.\footnote{In the final sample, we excluded one type 1 quasar that has an ETG-like morphology.}

To ensure the quality of measured $\lambda_{R_e}$, we did not use IFU data in which more than $25\%$ of spaxels within $R_e$ are excluded due to several conditions (median S/N $<5$, $\sigma<40$ km s$^{-1}$, $\vert V \vert\ge500$ km s$^{-1}$, or contamination from foreground stars, etc.) mentioned in Section \ref{sec:extract}. We also excluded IFU data in which the total numbers of spaxels within $R_e$ for the calculation of $\lambda_{R_e}$ are less than 45. Excluding such IFU data, the number of galaxies is 181.
\\

\subsection{Detection of Tidal Features}\label{sec:tf}
To detect tidal features, we performed visual inspection, as in \citet{YL2020}, on deep coadded images of the Stripe 82 from \citet{Jiang2014}.  Among the five bands of SDSS, we used $r$-band coadded images of which the surface brightness limit ($1\sigma$ of the background noise over a $1\arcsec\times1\arcsec$ region) is $\sim27$ mag arcsec$^{-2}$ and the $5\sigma$ detection limit of the aperture magnitude is 24.6.  We note that many studies still prefer visual inspection rather than quantitative methods \citep{Schawinski2010,Kaviraj2011,Atkinson2013,Hong2015,YL2020} as it can be difficult to detect faint irregular tidal features (or distinguish them from contaminations due to nearby sources) with the automatic and quantitative determination \citep{Kartaltepe2010,Miskolczi2011,Atkinson2013}. We examined each image to find tidal features such as tidal tails, shell structures, streams, and diffuse fans (see Figures \ref{fig:ex_1} and \ref{fig:ex_2}), adjusting the scale of pixel values. If needed, images were smoothed using Gaussian kernels to amplify signals of diffuse faint features. By examining the deep images, we found 14 ETGs that are too close to bright stars or bright large galaxies. The nearby pixels around such ETGs are highly contaminated by the bright sources, which makes it difficult to determine tidal features around the galaxies. Excluding them, the total number of galaxies in the final sample is 167. 

Deep coadded images of several ETGs that have tidal features are shown in the second rows of Figures \ref{fig:ex_1} and \ref{fig:ex_2}. Some ETGs without tidal features are shown in Figure \ref{fig:ex_3}, in which even deep images do not show any unusual features around ETGs. By contrast, in Figures \ref{fig:ex_1} and \ref{fig:ex_2}, tidal features that are not visible in the single-epoch color images are clearly observable around the ETGs in the deep images. We discovered that 41 ETGs have tidal features among 167 ETGs ($24.6\%$). 

The fraction of ETGs with tidal features found here ($24.6\%$) is far higher than that of \citet{Nair2010} who used normal SDSS images that are $\sim2$ mag shallower than the deep coadded images used here. They found that $\sim3\%$ of ETGs (E, ES0, and S0 types in their study) have tidal features such as shells and tails, and $6\%$ of ETGs have tidal or disturbed features. \citet{Hood2018} found that $18\%$ of bulged galaxies have tidal features using $r$-band images from the DECam Legacy Survey \citep{Dey2019}, whose surface brightness depth is $\sim0.4$ mag arcsec$^{-2}$ shallower than that of Stripe 82 coadded images. They also showed that the fraction can increase up to $\sim24\%$, which is similar to what we found here, if Stripe 82 coadded images were used. Extremely faint tidal features that we could not find here can be detected in images with great depth whose surface brightness limit is deeper than $\sim29$ mag arcsec$^{-2}$. For example, \citet{vanDokkum2005} found that $71\%$ of nearby ETGs have signatures of tidally disturbed features using such deep images.

By visually inspecting color images of SDSS, we classified ETGs that have prominent dust lanes. We found that 12 ETGs have clear dust lanes among 167 ETGs. The second, third, and fourth columns of Figure \ref{fig:ex_2} represent ETGs that have dust lanes and tidal features at the same time. As in \citet{YL2020}, these ETGs with dust lanes are treated separately in this study.

We compared our classifications of this study with those of \citet{Kaviraj2010} who classified ETGs with $M_r<-20.5$ and $z<0.05$ in the Stripe 82 region into several categories: relaxed ETGs, ETGs with tidal features, and ETGs with dust features. We found that 55 ETGs in our sample overlap with the ETG sample of \citet{Kaviraj2010}. In the overlapping 55 ETGs, we discovered tidal features in 20 ETGs, among which $90\%$ (18/20) were also determined as ETGs with tidal features in  \citet{Kaviraj2010}. By examining the deep images, we found that the two ETGs that \citet{Kaviraj2010} determined as relaxed ETGs (without tidal features) but that we classified as ETGs with tidal features have discernible tidal features: one has an obvious and prominent tidal feature and the other has a very faint shell structure. If we correct the definitely misclassified ETG in \citet{Kaviraj2010}, the percentage becomes $95\%$ (19/20). In the common 55 ETGs, \citet{Kaviraj2010} discovered tidal features in 19 ETGs, among which $94.7\%$ (18/19) were also detected as ETGs with tidal features in this study. The comparison suggests that our classifications for tidal features well agree with those from \citet{Kaviraj2010}.

In the case of dust lanes, we found five ETGs with prominent dust lanes, among which four ETGs ($80\%$) were classified as ETGs with dust features in \citet{Kaviraj2010}. We found that the ETG in disagreement has an obvious dust lane. Thus, the percentage becomes $100\%$ (5/5) if the misclassification in \citet{Kaviraj2010}  is corrected. That said, \citet{Kaviraj2010} detected dust features in six ETGs, among which four ETGs ($66.7\%$) were also classified as ETGs with dust lanes in this study. Although the two ETGs in disagreement have possible signs of faint dust features, we did not classify such ambiguous cases as ETGs with dust lanes in this study in order to define a clearly distinct ETG subpopulation.
\\

\begin{figure*}
\includegraphics[width=\linewidth]{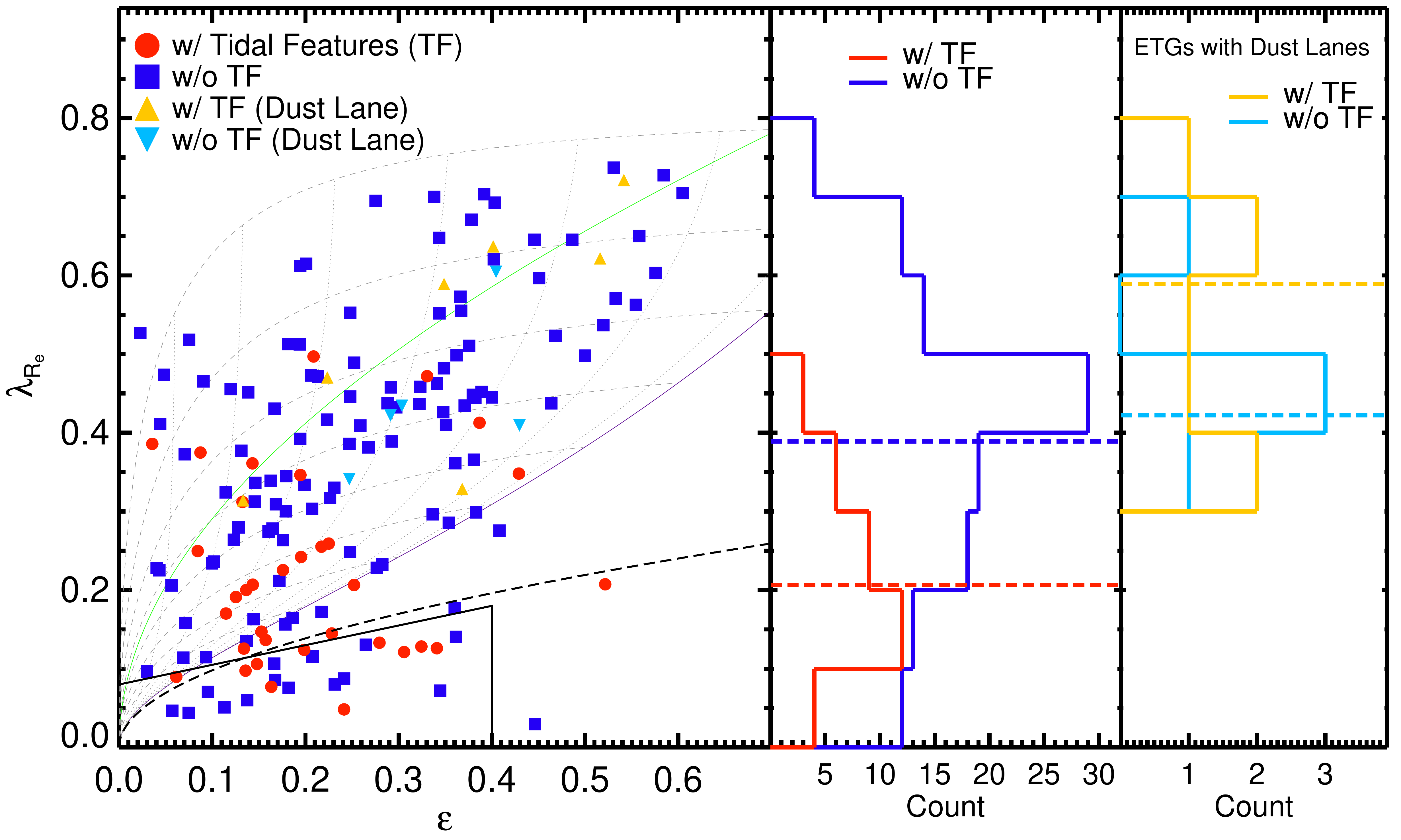}  
\centering
\caption{The left panel: the distribution of ETGs in the $\lambda_{R_e}$ versus $\varepsilon$ plane. ETGs are divided into four categories according to the existence of tidal features or dust lanes as shown in the legend. The green line denotes $\lambda_{R_e}$ for the edge-on isotropic rotator from \citet{Binney2005} with various intrinsic ellipticities \citep{Cappellari2016}. The purple line indicates $\lambda_{R_e}$ for the edge-on galaxies satisfying the anisotropy versus intrinsic flattening relation of Equation 11 in \citet{Cappellari2016}. The gray dotted lines represent variations of the purple line at different inclination angles with a step size of $0.1\degr$. The gray dashed lines show the traces of galaxies with a fixed intrinsic ellipticity moving across the diagram when the inclination angle changes. The black solid line and the black dashed line are the lines defining slow rotators in \citet{Cappellari2016} and \citet{Emsellem2011}, respectively (see Table \ref{tb:ft} for equations of the lines). The middle panel: $\lambda_{R_e}$ distributions for normal ETGs that do not have dust lanes. The right panel: $\lambda_{R_e}$ distributions for ETGs with dust lanes. The horizontal dashed lines in the middle and right panels indicate the median $\lambda_{R_e}$ for each ETG category.
\label{fig:main}}
\end{figure*}

\begin{deluxetable}{c ccc}
\tablecaption{Fraction of ETGs with tidal features \label{tb:ft}}
\tabletypesize{\scriptsize}
\tablehead{\colhead{Definition of ETG Population} & \colhead{$f_T$} & \colhead{$N_T$} & \colhead{$N_\mathrm{ETG}$} 
}
\startdata
Slow Rotators ($\lambda_{R_e}<0.08+\varepsilon/4$ and $\varepsilon<0.4$)\tablenotemark{a} & $0.417$ & $10$ & $24$\\  
Fast Rotators ($\lambda_{R_e}\ge0.08+\varepsilon/4$ or $\varepsilon\ge0.4$) & $0.183$ & $24$ & $131$\\ 
\hline 
Slow Rotators ($\lambda_{R_e}<0.31\sqrt{\varepsilon}$)\tablenotemark{b} & $0.407$ & $11$ & $27$\\  
Fast Rotators ($\lambda_{R_e}\ge0.31\sqrt{\varepsilon}$) & $0.180$ & $23$ & $128$\\ 
\hline 
$\lambda_{R_e}<0.25$ & $0.404$ & $23$ & $57$\\  
$\lambda_{R_e}\ge0.25$ & $0.112$ & $11$ & $98$\\ 
\enddata
\tablecomments{This table is for ETGs that do not have dust lanes. $f_T$: the fraction of ETGs with tidal features. $N_T$: the number of ETGs with tidal features. $N_\mathrm{ETG}$: the number of all ETGs (with and without tidal features). Thus, $f_T=N_T/N_\mathrm{ETG}$.}
\tablenotetext{a}{Equation 19 in \citet{Cappellari2016}}
\tablenotetext{b}{Equation 3 in \citet{Emsellem2011}}
\end{deluxetable}

\begin{figure}
\includegraphics[width=\linewidth]{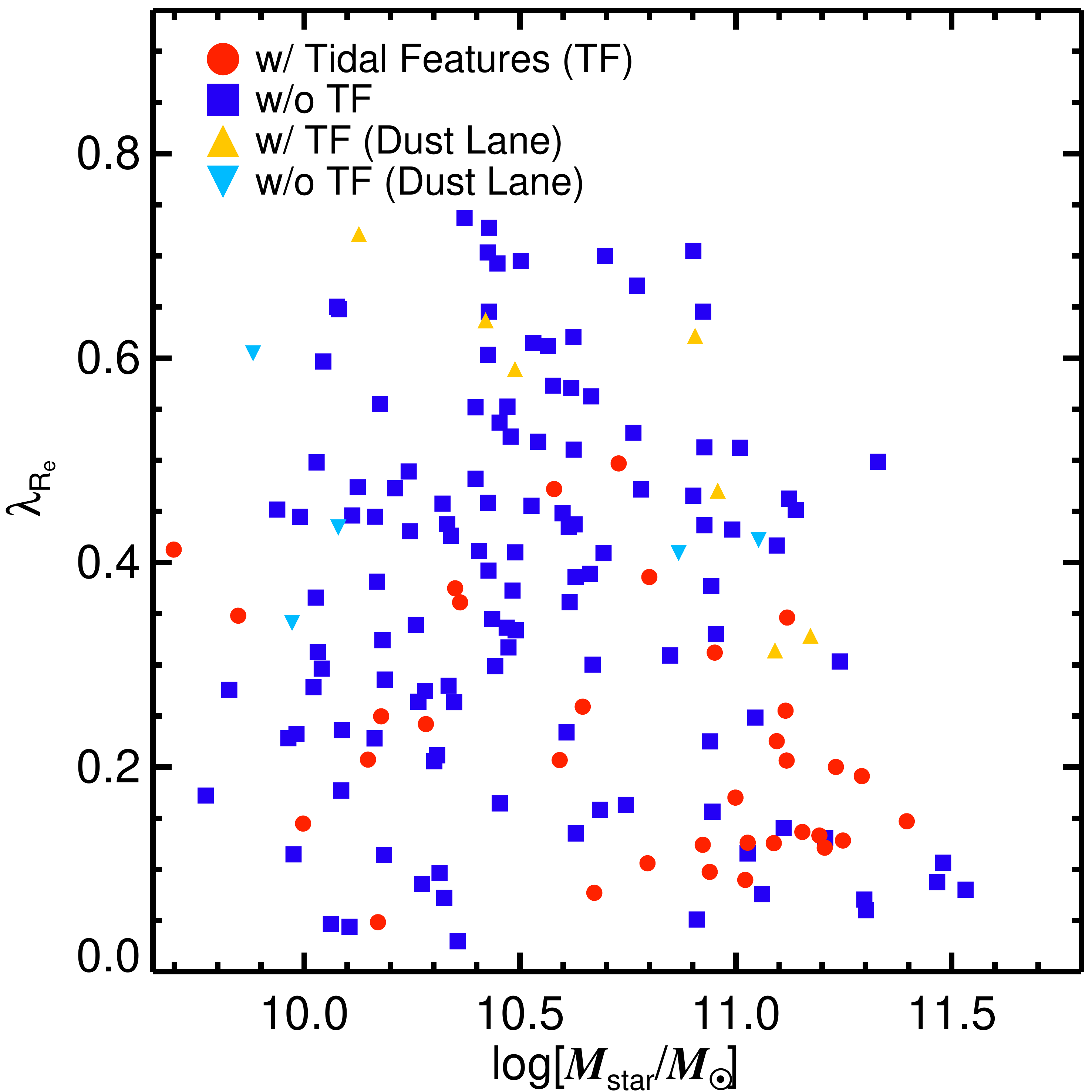}  
\centering
\caption{Distribution of ETGs in the $\lambda_{R_e}$ versus $\log M_\mathrm{star}$ plane. ETGs are divided into four categories according to the existence of tidal features or dust lanes as shown in the legend.
\label{fig:massl}}
\end{figure} 

\begin{figure*}
\includegraphics[width=\linewidth]{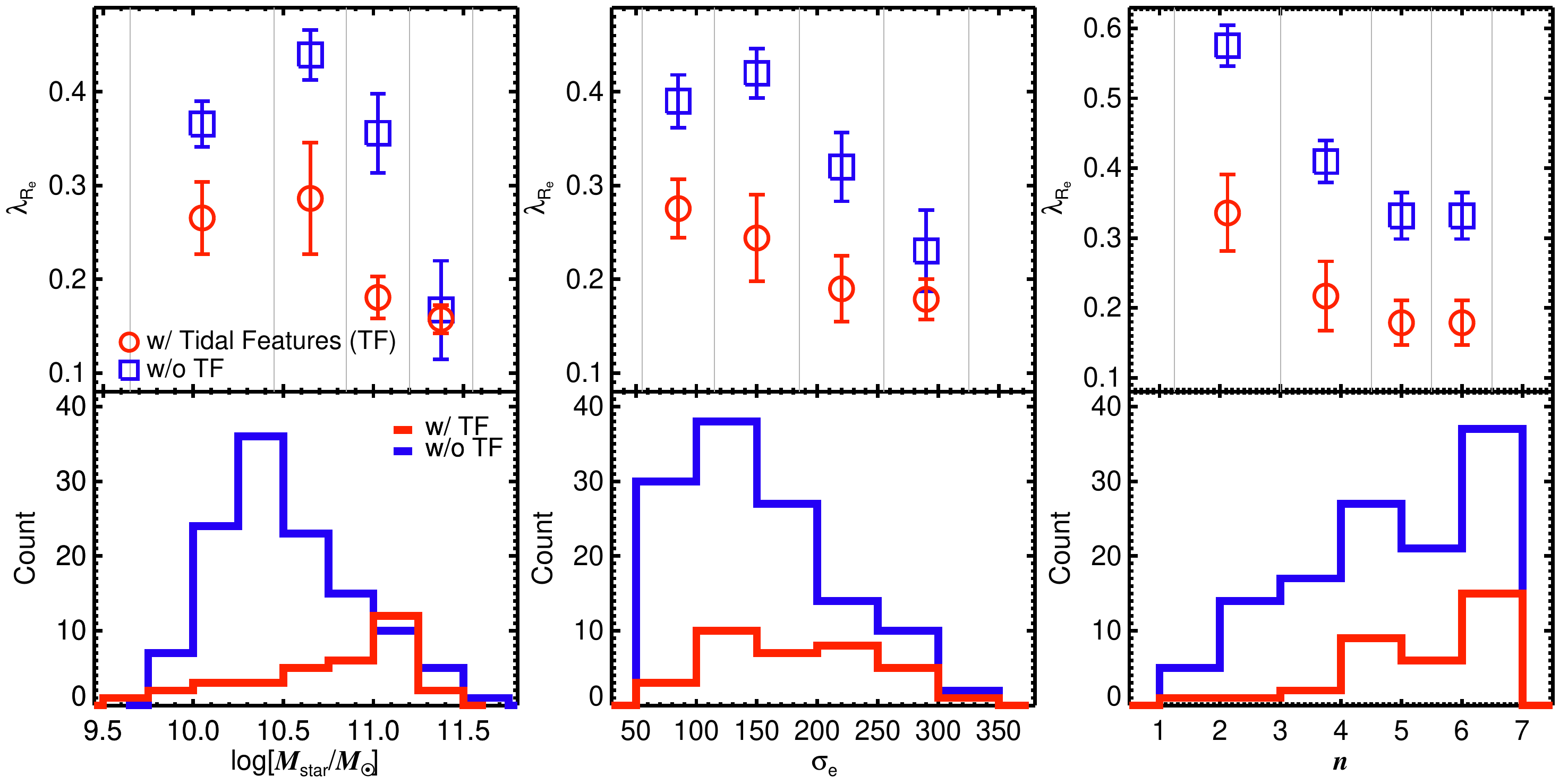}  
\centering
\caption{Mean $\lambda_{R_e}$ for ETGs with/without tidal features as a function of $\log M_\mathrm{star}$, $\sigma_e$, and $n$ (the upper panels). Also shown are histograms of $\log M_\mathrm{star}$, $\sigma_e$, and $n$ for the two ETG categories (the lower panels). In the upper panels, the error bars indicate standard deviations of the mean values from 1000 bootstrap resamplings, while the gray vertical lines denote the boundaries of the bins in which mean $\lambda_{R_e}$ values are computed. Normal ETGs that do not have dust lanes are used in this figure.
\label{fig:pdist}}
\end{figure*} 

\begin{figure}
\includegraphics[width=\linewidth]{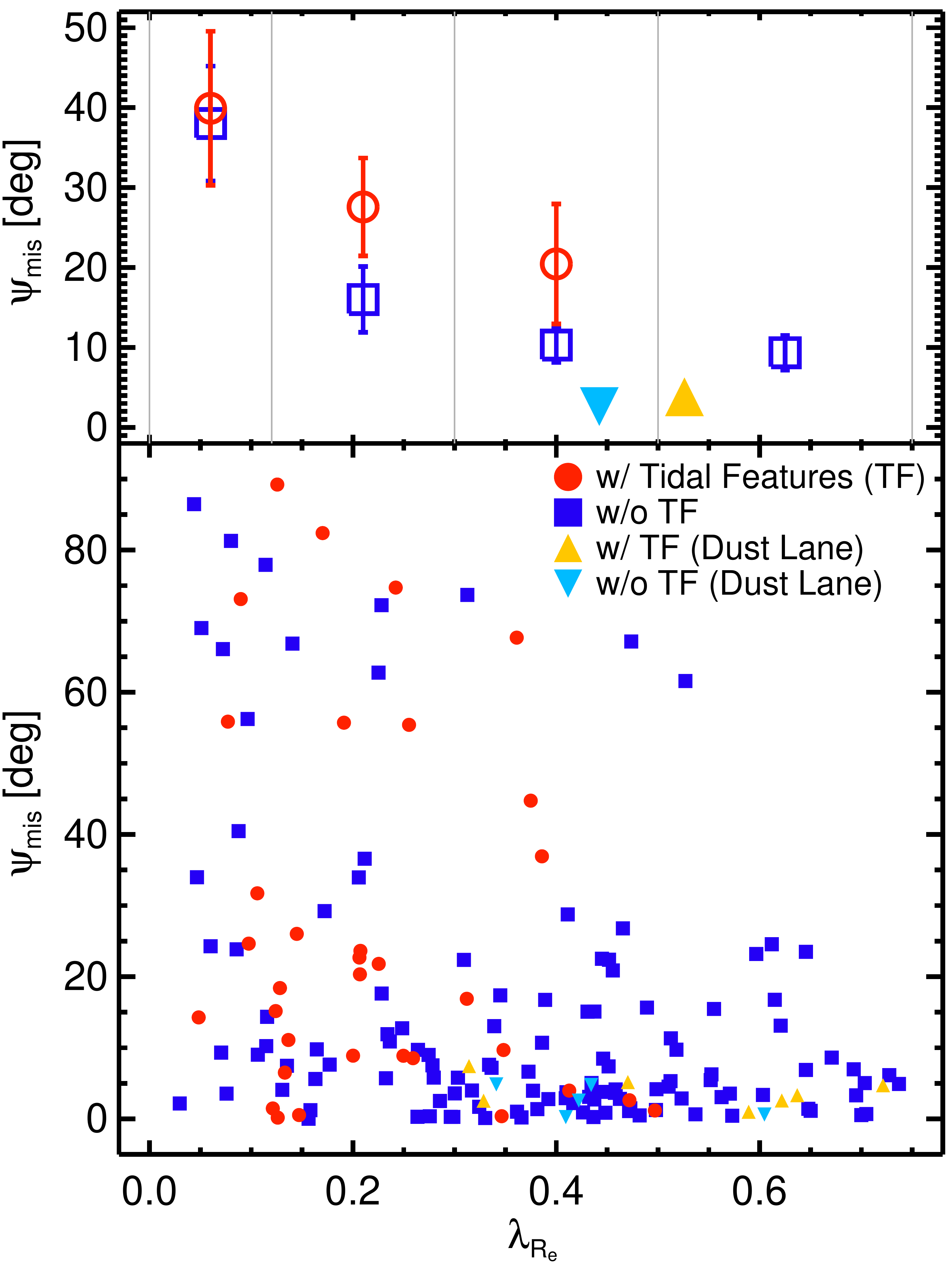}  
\centering
\caption{Distribution of ETGs in the $\psi_\mathrm{mis}$ versus $\lambda_{R_e}$ plane (the lower panels) and mean $\psi_\mathrm{mis}$ as a function of $\lambda_{R_e}$ (the upper panels). ETGs are divided into four categories according to the existence of tidal features or dust lanes as shown in the legend. In the upper panels, the error bars represent standard deviations of the mean values from 1000 bootstrap resamplings, while the gray vertical lines indicate the boundaries of the bins in which mean $\psi_\mathrm{mis}$ values are computed. In the upper panel, the positions of the triangles (ETGs with dust lanes) on the $x$-axis represent mean $\lambda_{R_e}$ values of the ETG categories.
\label{fig:padiff}}
\end{figure}

\section{Results}\label{sec:results}
Firstly, we describe results for normal ETGs that do not have dust lanes (hereafter, nETGs). The distribution of ETGs in the $\lambda_{R_e}$ versus $\varepsilon$ plane and its projected distributions for $\lambda_{R_e}$ in Figure \ref{fig:main} show that nETGs with tidal features have lower $\lambda_{R_e}$ than those without tidal features. The median $\lambda_{R_e}$ for nETGs with tidal features is $0.21\pm0.03$, while that of nETGs without tidal features is $0.39\pm0.03$.\footnote{The errors are the standard deviations of the median values from 1000 bootstrap resamplings.} We tested the significance of the difference between the two $\lambda_{R_e}$ distributions for nETGs (the middle panel in Figure \ref{fig:main}) using the Kolmogorov--Smirnov test and found that the probability ($0\le P \le1$) of the null hypothesis in which the two distributions are drawn from the same distribution is $1.9\times10^{-5}$, which means that the two distributions are significantly different.

We divided nETGs into slow rotators and fast rotators using two criteria in \citet{Emsellem2011} and \citet{Cappellari2016}. The black dashed line and the black solid line in Figure \ref{fig:main} indicate the lines dividing slow and fast rotators (see Table \ref{tb:ft} for equations for the lines). Then, we examined the fraction of nETGs with tidal features ($f_T$) in each kinematic class, which is shown in Table \ref{tb:ft}. We found that $f_T$ of slow rotators is more than twice as large as that of fast rotators: $18\%$ of fast rotators have tidal features, while $42\%$ of slow rotators have tidal features. When nETGs are divided by $\lambda_{R_e}=0.25$, only $11\%$ of nETGs with $\lambda_{R_e}\ge0.25$ have tidal features, whereas $40\%$ of nETGs with $\lambda_{R_e}<0.25$ show tidal features.

We also investigated how $\lambda_{R_e}$ values in the different nETG categories depend on $M_\mathrm{star}$, galaxy structure, and effective velocity dispersion within $R_e$ ($\sigma_e$). For galaxy structure, we used the S{\'e}rsic index ($n$) from the NSA catalog. Dispersion $\sigma_e$ was calculated by Equation 3 in \citet{Graham2018}. The results are shown in Figures \ref{fig:massl} and \ref{fig:pdist}. Figure \ref{fig:massl} displays the distribution of ETGs in the $\lambda_{R_e}$ versus $\log M_\mathrm{star}$ plane. Figure \ref{fig:pdist} shows the mean $\lambda_{R_e}$ for nETGs with/without tidal features as a function of $\log M_\mathrm{star}$, $\sigma_e$, and $n$. Also shown in Figure \ref{fig:pdist} are histograms of $\log M_\mathrm{star}$, $\sigma_e$, and $n$ for the two ETG categories. 

The two figures demonstrate that the mean $\lambda_{R_e}$ for nETGs without tidal features is $\sim0.4$ at $\log(M_\mathrm{star}/M_\odot)<11.2$ and falls to $0.17\pm0.05$ at the mass range of $\log(M_\mathrm{star}/M_\odot)\ge11.2$.\footnote{Previous studies discovered that the fraction of slow rotators increases at $\log(M_\mathrm{star}/M_\odot)\gtrsim11.2$ \citep{Emsellem2007,Graham2018}.} By contrast, the mean $\lambda_{R_e}$ values for nETGs with tidal features are $0.18$--$0.29$ at $\log(M_\mathrm{star}/M_\odot)<11.2$, which are lower than those of nETGs without tidal features in the same mass bins. The mean $\lambda_{R_e}$ for nETGs with tidal features at $\log(M_\mathrm{star}/M_\odot)\ge11.2$ is $0.16\pm0.01$, so that nETGs in both the categories have similar mean $\lambda_{R_e}$ values at $\log(M_\mathrm{star}/M_\odot)\ge11.2$. 

The trend in $\lambda_{R_e}$ as a function of $\sigma_e$ shown in the middle panel of Figure \ref{fig:pdist} is similar to the case of $M_\mathrm{star}$ described above in the sense that $\lambda_{R_e}$ is lower at higher $\sigma_e$ and nETGs with tidal features have lower $\lambda_{R_e}$ than the counterparts, especially at $\sigma_e<255$ km s$^{-1}$.\footnote{See also \citet{Graham2018} for the relation between $\lambda_{R_e}$ and $\sigma_e$.} 

The right panel of Figure \ref{fig:pdist} shows that as $n$ rises, the mean $\lambda_{R_e}$ for nETGs without tidal features decreases from 0.58 to 0.33, while that of nETGs with tidal features changes from 0.34 to 0.18. Thus, the mean $\lambda_{R_e}$ is lower at higher $n$ and nETGs with tidal features have lower $\lambda_{R_e}$ than the counterparts in all of the $n$ bins. We note that the trends for $\lambda_{R_e}$ shown in the middle and right panels of Figure \ref{fig:pdist} are the reflection of what the left panel of the figure displays ($\lambda_{R_e}$ as a function of $M_\mathrm{star}$), since $\sigma_e$ and $n$ are well correlated with $M_\mathrm{star}$.

The histograms in Figure \ref{fig:pdist} tell us that tidal features are more frequent in more massive ETGs and ETGs with more concentrated light distribution (higher $n$). This is consistent with the results in \citet{YL2020}.

We present the results for $\psi_\mathrm{mis}$ in Figure \ref{fig:padiff}, which displays the distribution of ETGs in the $\psi_\mathrm{mis}$ versus $\lambda_{R_e}$ plane and the mean $\psi_\mathrm{mis}$ as a function of $\lambda_{R_e}$. In the figure, we found that $\psi_\mathrm{mis}$ depends on $\lambda_{R_e}$ in such a way that ETGs with lower $\lambda_{R_e}$ have larger $\psi_\mathrm{mis}$, which is consistent with previous results \citep{Emsellem2007,Oh2016,Graham2018}. The decreasing trend in $\psi_\mathrm{mis}$ as a function of $\lambda_{R_e}$ is more severe in nETGs without tidal features than in the counterparts with tidal features. The mean $\psi_\mathrm{mis}$ at $\lambda_{R_e}<0.12$ is $\sim40\degr$ for both the nETG categories. However, the mean $\psi_\mathrm{mis}$ falls to $9\degr$ at $\lambda_{R_e}\ge0.3$ for nETGs without tidal features, while it decreases to $20\degr$ at $\lambda_{R_e}\ge0.3$ for nETGs that have tidal features. Thus, $\psi_\mathrm{mis}$ is larger in nETGs with tidal features\footnote{\citet{Krajnovic2011} and \citet{Oh2016} found that ETGs with disturbed features or evidence for recent interactions show large kinematic misalignments.} except at the lowest $\lambda_{R_e}$ bin. 

We note that low $\lambda_{R_e}$ values in nETGs with tidal features are not due to the large $\psi_\mathrm{mis}$. Even if we calculate $\lambda_{R_e}$ of nETGs with tidal features based on ellipses aligned with kinematic position angles, $\lambda_{R_e}$ values change by only less than $\sim0.01$.

Although the number is small, ETGs that have dust lanes (hereafter, dETGs) show different trends compared to nETGs. As shown in Figures \ref{fig:main} and \ref{fig:massl}, dETGs are all fast rotators. Moreover, dETGs with tidal features have higher $\lambda_{R_e}$ than those without tidal features, which is a reverse trend to what nETGs show. The median $\lambda_{R_e}$ for dETGs without tidal features is $0.42\pm0.05$, which is comparable to that of nETGs without tidal features. That said, the median $\lambda_{R_e}$ for dETGs with tidal features is $0.59\pm0.10$, which is the highest value among those of the four ETG categories. Considering the fact that the median $n$ for dETGs with tidal features is $4.0$ (the mean $n$ is $3.8$), the high $\lambda_{R_e}$ in dETGs with tidal features is not merely attributable to the simple bias in their $n$ values (see Figure \ref{fig:pdist}).

We also found that $\psi_\mathrm{kine}$ is well aligned with $\psi_\mathrm{phot}$ (small $\psi_\mathrm{mis}$) in dETGs regardless of the existence of tidal features as shown in Figure \ref{fig:padiff}. Their $\psi_\mathrm{mis}$ values are all smaller than $7.5\degr$.
\\

\section{Discussion}\label{sec:discuss}

Galaxy mergers and close encounters not only affect color and internal structures \citep{Schweizer1992,Tal2009,Schawinski2010,Kaviraj2011,Hong2015,YL2020}, but also have an influence on stellar kinematics of merger remnants \citep{Hwang2015,Hwang2018,Hwang2021}. Kinematic properties of merger remnants depend on a wide diversity of factors such as gas fractions, mass ratios of progenitors, merger orbits, relative orientations of progenitors, etc \citep{Cox2006,Jesseit2009,Hoffman2010,Bois2011,Naab2014,Penoyre2017}. Therefore, ETGs are bound to have varied kinematic characteristics according to their different merger/formation histories. The wide ranges of $\lambda_{R_e}$ and $\psi_\mathrm{mis}$ for ETGs shown in Section \ref{sec:results} are likely to be the consequence of various merger/formation histories. 

ETGs with tidal features are distinguished from those without tidal features in that they have experienced recent mergers or close interactions within a few gigayears \citep{Ji2014,Mancillas2019,YL2020}. One notable (and average) difference between recent mergers and mergers that occurred so long ago that no traces are left behind is gas fractions in galaxy mergers. While there is no obvious reason for other factors in mergers such as merger orbits and relative orientations of progenitors to be different in different redshifts, it is known that (cold) gas fractions in galaxies (progenitors) are lower in lower redshifts \citep{Erb2006,Daddi2010,Tacconi2010,Geach2011,Carilli2013,Morokuma2015,Yoon2019} as galaxies gradually consume gas in various ways in the history of the universe. 

In this sense, nETGs with tidal features are expected to have experienced relatively more dissipationless (dry) mergers than nETGs without tidal features. On average, such mergers with relatively lower gas fractions can make remnants that are more slowly rotating and have larger kinematic misalignments \citep{Barnes1992,Cox2006,Emsellem2007,Hoffman2010,Lagos2017}. Thus, this picture is consistent with our results in that nETGs with tidal features have statistically lower $\lambda_{R_e}$ and larger $\psi_\mathrm{mis}$ than nETGs without tidal features. 

Fast rotating nETGs with low S{\'e}rsic indices of $n\lesssim2.5$ that do not have tidal features are possible candidates for passively evolved versions of spiral galaxies without experiences of significant merger activities. The existence of such galaxies in our ETG sample may partly contribute to the difference in $\lambda_{R_e}$ and  $\psi_\mathrm{mis}$ between the two ETG categories. 

Our results show that more massive ETGs have on average lower $\lambda_{R_e}$. This is related to the fact that the fraction of stellar mass in an ETG that was accreted from mergers is significantly larger in more massive ETGs, as found in a number of recent observational \citep{Oyarzun2019,Davison2021} and theoretical \citep{Rodriguez2016,Davison2020} studies, since consecutive mergers in the past can slow down rotating systems \citep{Bournaud2007,Qu2010,Khochfar2011,Choi2017}. Likewise, it is possible that most very massive ETGs with $\log(M_\mathrm{star}/M_\odot)\sim11.4$ that do not have tidal features already turned into systems as slow as ETGs with tidal features a long time ago, due to numerous gas-poor mergers that they have experienced in their formation histories \citep{Penoyre2017,Yoon2017}.

A possible origin of dust lanes in ETGs is a gas-rich merger process \citep{Oosterloo2002,Kaviraj2012,Shabala2012}. Though gas-rich mergers are more frequent at higher redshifts, they can still occur at low redshifts and trigger dusty starbursts \citep{Chen2010}. A gas-rich merger induces gas inflows into the inner region of the merger remnant due to loss of angular momentum through radiation and tidal torques during the merger process, and then the gas can form young stars \citep{Hernquist1989,Barnes1991,Barnes1996,Hopkins2008a} that are rotating fast with angular momentum aligned with the minor axis of the remnant \citep{SH2005,Cox2006,Robertson2006,Hopkins2009,Hoffman2010}. Since gas-rich mergers are able to cause baryonic matter distributions to be centrally concentrated in merger remnants, old stars (stars that already existed before the merger) in the gas-rich merger remnant can also rotate relatively faster than those in a dry merger remnant due to the steep gravitational potential \citep{Cox2006,Hoffman2010}. In this way, gas-rich mergers, on average, can make ETGs that rotate fast with small kinematic misalignments compared to dry mergers \citep{Cox2006,Robertson2006,Emsellem2007,Hoffman2010,Lagos2017,Penoyre2017}, although there are wide variations depending on other factors such as merger orbits or mass ratios and orientations of progenitors.

Considering our results that dETGs with tidal features have statistically high $\lambda_{R_e}$ and small $\psi_\mathrm{mis}$ compared with other ETGs, a recent merger between gas-rich progenitors (at least more gas-rich than the progenitors of nETGs with tidal features) can be a reasonable origin of ETGs with dust lanes that have tidal features. However, a larger deep imaging sample with a sufficient number of ETGs having dust lanes is necessary to discuss their properties and origin more clearly.

Our results suggest that kinematic properties within a few kiloparsecs of galaxy centers correlate with the existence of tidal features that span tens of kiloparsecs around galaxies. This implies that galaxy mergers that can leave large-scale tidal features that are prominent enough to be detectable in the coadded images play a key role in the evolution of stellar kinematics of galaxy centers.
\\

\section{Summary}\label{sec:summary}

We examined the impact of recent galaxy mergers on stellar kinematics of ETGs such as stellar angular momentum within the half-light radius $\lambda_{R_e}$ and kinematic misalignment $\psi_\mathrm{mis}$, using MaNGA data that are in the Stripe 82 region of SDSS. To mitigate the seeing effect in stellar kinematics, deconvolution is conducted on the MaNGA IFU data using the LR algorithm. Stellar velocities and velocity dispersions were extracted using the pPXF code that performs full spectrum fitting on galaxy spectra. We detected tidal features, which are direct evidence for recent mergers, by visually inspecting deep coadded images of the Stripe 82 region that are $\sim2$ mag deeper than single-epoch images of SDSS. The number of ETGs in the final sample, which have $z <0.055$, is 167. We categorized ETGs into four groups according to the existence of tidal features or dust lanes and investigated their stellar kinematics. The main results of this study are as follows.

\begin{itemize}
\item{Results for normal ETGs that do not have dust lanes:}
\begin{enumerate}
\item ETGs with tidal features have lower $\lambda_{R_e}$ than those without tidal features: the median $\lambda_{R_e}$ for ETGs with tidal features is $0.21$, while that of ETGs without tidal features is $0.39$.

\item The fraction of ETGs with tidal features in slow rotators is more than twice as large as that in fast rotators ($42\%$ versus $18\%$).

\item In all mass and S{\'e}rsic index ranges except for the most massive bin, ETGs with tidal features have lower $\lambda_{R_e}$ than those without tidal features.

\item Misalignment $\psi_\mathrm{mis}$ is larger in ETGs with tidal features than those without tidal features (at a given $\lambda_{R_e}$ when $\lambda_{R_e}\gtrsim0.1$): the mean $\psi_\mathrm{mis}$ for ETGs with tidal features is $27.5\degr$, while that of ETGs without tidal features is $15.0\degr$.
\end{enumerate}

\item{Results for ETGs that have dust lanes:}
\begin{enumerate}
\item ETGs with dust lanes are fast rotators.

\item ETGs with both dust lanes and tidal features have the highest $\lambda_{R_e}$ (the median $\lambda_{R_e}$ is $0.59$) among all ETG categories. 

\item ETGs with dust lanes have small $\psi_\mathrm{mis}$ regardless of the existence of tidal features: their $\psi_\mathrm{mis}$ values are all smaller than $7.5\degr$. 
\end{enumerate}
\end{itemize}

Our results can be explained if mergers with higher (lower) gas fractions, on average, make remnants that rotate faster (more slowly) and have smaller (larger) kinematic misalignments. The results of this study provide observational evidence that galaxy mergers significantly affect stellar kinematics of ETGs. Future studies using larger and deeper imaging data can verify further details about the impact of galaxy mergers on stellar kinematics of galaxies.
\\

\clearpage

\begin{acknowledgments}
This work was supported by a KIAS Individual Grant PG076302 at the Korea Institute for Advanced Study.
Funding for the Sloan Digital Sky 
Survey IV has been provided by the 
Alfred P. Sloan Foundation, the U.S. 
Department of Energy Office of 
Science, and the Participating 
Institutions. 

SDSS-IV acknowledges support and 
resources from the Center for High 
Performance Computing  at the 
University of Utah. The SDSS 
website is www.sdss.org.

SDSS-IV is managed by the 
Astrophysical Research Consortium 
for the Participating Institutions 
of the SDSS Collaboration including 
the Brazilian Participation Group, 
the Carnegie Institution for Science, 
Carnegie Mellon University, Center for 
Astrophysics | Harvard \& 
Smithsonian, the Chilean Participation 
Group, the French Participation Group, 
Instituto de Astrof\'isica de 
Canarias, The Johns Hopkins 
University, Kavli Institute for the 
Physics and Mathematics of the 
Universe (IPMU) / University of 
Tokyo, the Korean Participation Group, 
Lawrence Berkeley National Laboratory, 
Leibniz Institut f\"ur Astrophysik 
Potsdam (AIP),  Max-Planck-Institut 
f\"ur Astronomie (MPIA Heidelberg), 
Max-Planck-Institut f\"ur 
Astrophysik (MPA Garching), 
Max-Planck-Institut f\"ur 
Extraterrestrische Physik (MPE), 
National Astronomical Observatories of 
China, New Mexico State University, 
New York University, University of 
Notre Dame, Observat\'ario 
Nacional / MCTI, The Ohio State 
University, Pennsylvania State 
University, Shanghai 
Astronomical Observatory, United 
Kingdom Participation Group, 
Universidad Nacional Aut\'onoma 
de M\'exico, University of Arizona, 
University of Colorado Boulder, 
University of Oxford, University of 
Portsmouth, University of Utah, 
University of Virginia, University 
of Washington, University of 
Wisconsin, Vanderbilt University, 
and Yale University.
\end{acknowledgments}

\end{document}